\documentclass{aa}
\usepackage{graphicx}
\usepackage{txfonts}

\begin{document}

   \title{Metallicity of M dwarfs}
   \subtitle{I. A photometric calibration and the impact on the mass-luminosity 
relation\\ at the bottom of the main sequence
\thanks{Based on observations obtained with the ELODIE 
           spectrograph on the 1.93~m telescope of the Observatoire de 
           Haute Provence, France}
}

   \author{X. Bonfils
          \inst{1,2}
          \and
          X. Delfosse\inst{1}
          \and
          S. Udry\inst{2}
          \and
          N.C. Santos\inst{2,3}
          \and
          T. Forveille\inst{1,4}
          \and
          D. S\'egransan\inst{2}
          }
   \offprints{X. Bonfils, \\
   \email{Xavier.Bonfils@obs.ujf-grenoble.fr}}
   \institute{Laboratoire d'Astrophysique, Observatoire de Grenoble, 
        BP 53, F-38041 Grenoble, C\'edex 9, France
     \and
     Observatoire de Gen\`eve, 51 ch. des Maillettes, 1290 Sauverny, 
        Switzerland
     \and
     Centro de Astronomia e Astrofisica da Universidade de Lisboa, 
        Observat\'orio Astr\'onomico de Lisboa, Tapada de Ajuda, 1349-018 
        Lisboa, Portugal
     \and
     Canada-France-Hawaii Telescope Corporation, 65-1238 Mamalahoa Highway, 
        Kamuela, HI~96743, Hawaii, USA
   }

   \date{Received March 14, 2005; accepted May 19, 2005}

   \abstract{We obtained high resolution ELODIE and CORALIE spectra for both 
components of 20 wide visual binaries composed of an F-, G- or K-dwarf primary
 and an M-dwarf secondary. We analyse the well-understood spectra of the 
primaries to determine metallicities ([Fe/H]) for these 20 systems, and hence 
for their M dwarf components. We pool these metallicities with determinations 
from the literature to obtain a precise ($\pm$ 0.2 dex) photometric 
calibration of M dwarf metallicities. This calibration represents a 
breakthrough in a field where discussions have had to remain largely 
qualitative, and it helps us demonstrate that metallicity explains most 
of the large dispersion in the empirical V-band mass-luminosity relation. 
We examine the metallicity of the two known M-dwarf planet-host stars, 
\object{Gl 876} ($+$0.02 dex) and \object{Gl 436} ($-$0.03 dex), in the 
context of preferential planet formation around metal-rich stars. We 
finally determine the metallicity of the 47 brightest single M dwarfs 
in a volume-limited sample, and compare the metallicity distributions 
of solar-type and M-dwarf stars in the solar neighbourhood.
   
     }

   \maketitle

\section{Introduction}
\label{sect:intro}
The very low mass M dwarfs are small, cool and faint, but they dominate the 
Galaxy by number ($\sim$50-70\%), and even by total mass 
($\sim$70\%)(Chabrier \cite{chabrier03}). Any 
realistic model of the Galaxy therefore needs an excellent description of 
this faint component. Over the last decade, stellar models of very low mass 
stars have made great strides, but they still have to use some incomplete or 
approximate input physics (Chabrier \& Baraffe \cite{chabrier00}). 
Descriptions of these stars therefore need a strong empirical basis, or 
validation.

In S\'egransan et al. (\cite{segransan03}) and Delfosse et al. 
({\cite{delfosse00}, hereafter DFS00), we have validated 
the model predictions for radii and luminosities. The empirical radii 
match the models very well, and have no dispersion beyond the measurement 
errors. The infrared \emph{mass-luminosity} (hereafter M/L) relations also 
have negligible dispersion, and similarly agree with model predictions. 
The V-band M/L relation, in contrast, has a large (${\sim}{\pm}1~mag$) 
intrinsic scatter. In DFS00 we suggested that metallicity might explain 
most of this intrinsic dispersion, but for lack of quantitative metallicity 
estimates we could not pursue this suggestion.
  
\begin{table*}
	\caption[]{Observed visual binaries with an M-dwarf secondary.}
	\label{table:targets}
\begin{tabular}{          l r@{:}c@{:}lr@{:}c@{:}l l l  l r@{:}c@{:}lr@{:}c@{:}l l l}
		\hline
\multicolumn{9}{c}{Primaries}&\multicolumn{9}{c}{Secondaries}\\
Name      &\multicolumn{3}{c}{$\alpha$ (2000)} & \multicolumn{3}{c}{$\delta$ (2000)} &m$_V$   &Sp. Typ.
&Name     &\multicolumn{3}{c}{$\alpha$ (2000)} & \multicolumn{3}{c}{$\delta$ (2000)} &m$_V$   &Sp. Typ.\\
		\hline
GJ\,1021	         &00&45&45.593	&$-$47&33&07.15    &5.80        &G1V    &CD-48\,176\,B	&00&45&43.5 	&$-$47&33&00 	&13.5 	&M\\
Gl\,34\,A     	&00&49&06.291	&$+$57&48&54.67	&3.44	&F9V     &Gl\,34\,B    	         &00&49&06.5 	&$+$57&48&55   	&7.51  	&K7\\
Gl\,53.1\,A  	&01&07&37.872	&$+$22&57&17.91	&8.41	&K4V     &Gl\,53.1\,B         &01&07&37.7   	&$+$22&57&18    &13.60  &M3\\
Gl\,81.1A           &01&57&09.607        &$-$10&14&32.75   &6.42        &G5       &Gl\,81.1\.B         &01&57&11.1     &$-$10&14&53    &11.21  &K7\\
Gl\,105\,A  	&02&36&04.894	&$+$06&53&12.73	&5.79	&K3V    &Gl\,105\,B          &02&36&15.3     &$+$06&52&19    &11.68  &M4\\
Gl\,107\,A  	&02&44&11.986	&$+$49&13&42.41	&4.10	&F7V    &Gl\,107\,B          &02&44&11.8   	&$+$49&13&43    &10.06  &M1.5\\
GJ\,3194\,A  	&03&04&09.636	&$+$61&42&20.99	&6.64	&G4V    &GJ\,3195\,B      &03&04&43.6     &$+$61&44&08    &12.5   &M3\\
Gl\,166\,A  	&04&15&16.320	&$-$07&39&10.33	&4.43	&K1V    &Gl\,166\,C         &04&15&18.5   	&$-$07&39&07    &11.17  &M4.5\\
Gl\,211	  	&05&41&20.336	&$+$53&28&51.81	&6.21	&K1       &Gl\,212     	       &05&41&30.7	&$+$53&29&23   	&9.80   &M0.5\\
Gl\,231.1\,A  	&06&17&16.138	&$+$05&06&00.40	&5.70	&F9V    &Gl\,231.1\,B      &06&17&11.0  	&$+$05&07&06    &13.42 	&M3.5\\
Gl\,250\,A  	&06&52&18.050	&$-$05&10&25.37	&6.59	&K3      &Gl\,250\,B          &06&52&18.1   	&$-$05&11&26    &10.09  &M2\\
Gl\,297.2\,A  	&08&10&39.826	&$-$13&47&57.15	&5.53	&F7       &Gl\,297.2\,B  &08&10&34.0   	&$-$13&48&48    &11.80  &M2\\
Gl\,324\,A  	&08&52&35.811	&$+$28&19&50.95	&5.53	&G8      &Gl\,324\,B    &08&52&40.8   	&$+$28&18&59    &13.14  &M3.5\\
Gl\,505\,A  	&13&16&51.052	&$+$17&01&01.86	&6.69	&K2      &Gl\,505\,B    &13&16&51.7 	&$+$17&00&56   	&9.6  &M0.5\\
Gl\,544\,A  	&14&19&34.864	&$-$05&09&04.30	&7.58	&K1      &Gl\,544\,B    &14&19&35.0  	&$-$05&09&08    &14.1   &M6\\
NLTT\,45789    &18&00&38.894        &$+$29&34&18.91  &7.07        &G2V    &NLTT\,45791   &18&00&45.4     &$+$29&33&57    &13.1   &M\\
Gl\,768.1\,A  	&19&51&01.643	&$+$10&24&56.62   &5.12	&F8V    &Gl\,768.1\,B  &19&51&01.1   	&$+$10&24&43    &13.1  	&M3.5\\
Gl\,783.2\,A  	&20&11&06.074	&$+$16&11&16.80	&7.34	&K1V    &Gl\,783.2\,B  &20&11&13.4   	&$+$16&11&07    &13.94  &M4\\
Gl\,797\,A  	&20&40&45.141	&$+$19&56&07.93	&6.43	&G5V   &Gl\,797\,B    &20&40&44.4   	&$+$19&53&59    &11.88  &M2.5\\
Gl\,806.1\,A  	&20&46&12.683	&$+$33&58&12.92	&2.48	&K0III   &Gl\,806.1\,B  &20&46&12.7  	&$+$33&58&12    &13.4   &M3\\
Gl\,872\,A  	&22&46&41.581	&$+$12&10&22.40	&4.20	&F7V    &Gl\,872\,B   &22&46&41.6	&$+$12&10&20    &11.7  	&M1\\
		\hline
  	\end{tabular}
\end{table*}

M-dwarf metallicities have also become relevant in the context of planet 
formation around very low mass stars. One robust result of the exoplanet 
searches is that G and K stars which host planets are on average more 
metal-rich than the bulk of the solar neighbourhood population 
(Gonzalez \cite{gonzalez97}; Santos et al. \cite{santos01}, 
\cite{santos03}, \cite{santos04}). A leading explanation for this 
observation is that the disks of metal-rich stars contain larger 
amounts of refractory dust, and that more massive dust disks are 
much more likely to form planets. This has a clear bearing on planets 
around M dwarfs, since these low mass stars are likely to have smaller 
disks than solar-type stars of the same metallicity. Assuming that 
protostellar disk mass scales with stellar mass and within the 
core-accretion scenario, Laughlin et al. 
(\cite{laughlin04}) and Ida \& Lin (\cite{ida05}) show that
formation of Jupiter-mass planets is seriously inhibited around the 
less massive M dwarfs (M$_\star < 0.4 $M$_\odot$). To date, the only 
two M dwarfs known to host planets are \object{Gl 876} (Delfosse et al. 
\cite{delfosse98}; Marcy et al. \cite{marcy98}, \cite{marcy01}) and 
\object{Gl 436} (Butler et al. \cite{butler04}), but a number of 
ongoing surveys are looking for more (e.g. Bonfils et al. \cite{bonfils04}; 
Endl et al. \cite{endl03}; Kuerster et al. \cite{kuerster03}; 
Wright et al. \cite{wright04}). They will bring new constraints 
on the frequency of planets as a function of stellar mass, and 
metallicity will be one of the key parameters in the comparison with 
solar-type targets.
  
Measuring M-dwarf metallicities from their spectra is unfortunately
difficult. As the spectral subtype increases, the atmospheres of these 
cool stars ($\sim3800$~K (M0)~$>$~T~$_{eff}>\sim2100$~K (M9)) contain 
increasingly abundant diatomic and triatomic molecules 
(\element{TiO}, \element{VO}, H$_2$O,
\element{CO}, \element{FeH}, \element{CrH}...), 
which spectroscopically defines the M class. These components have complex
and extensive absorption band structures, which eventually leave no continuum
point in the spectrum. In a late-M dwarf, the local pseudo-continuum estimated
from a high resolution spectrum is defined by a forest a weak lines, and often
underestimates the true continuum by a factor of a few. The ``line-by-line'' 
spectroscopic analysis used for hotter stars therefore becomes impossible 
for late-M dwarfs, and a full spectral synthesis must be used. Besides the 
practical complexities of that approach, the atmospheric models do
not yet reproduce the details of high resolution spectra (mostly due
to limitations of their molecular opacity databases). This therefore
leaves some doubt about the reliability of the resulting metallicities.
Here we instead observe visual binaries that contain both an M-dwarf 
and a solar-type star. They presumably share a common metallicity  
that reflects the composition of their parent molecular cloud, and we use 
the much better understood spectrum of the solar-type star to infer the 
metallicity of the M dwarf.

In Sect. \ref{sect:previous estimates} we review the limited literature 
on observational M-dwarf metallicities. Sect. \ref{sect:data analysis} 
briefly describes the binary sample, the observations, and our analysis
of the primary star spectra. Sect. \ref{sect:calibration} describes the derivation of a photometric 
metallicity estimator for very low-mass stars. Sect. \ref{sect:mlfunction}
re-examines the dispersion of the V-band M/L relation in the light 
of the new metallicities and proposes a more precise 
{\it mass-metallicity-luminosity} relation for very low-mass stars.
In Sect. \ref{sect:planets} we apply the metallicity estimator 
to the two known M-dwarf planet-host stars. 
Sect. \ref{sect:9pc} lists estimated metallicities for
a volume-limited sample of northern M dwarfs, and compares its
metallicity distribution with that of nearby solar-type stars.

\section{Previous metallicity estimates of M dwarfs}
\label{sect:previous estimates}
The first attempt to measure metallicities for M dwarfs was
by Mould (\cite{mould76}, \cite{mould78}) who performed
line-by-line analyses of atomic lines in intermediate resolution 
near-IR spectra of a few stars, using model atmospheres available then. 
Jones et al. (\cite{jones96}) used a similar approach, but with the 
benefit of atmospheric structures from an early version of the modern 
PHOENIX code (Allard \& Hauschildt, \cite{allard95}). Gizis 
(\cite{gizis97}) matched low resolution optical spectra 
to synthetic spectra from the same Allard \& Hauschildt 
(\cite{allard95}) models, and derived relatively crude 
metallicities that allowed them to classify M dwarfs into
3 broad categories (dwarfs, subdwarfs and  extreme subdwarfs). 
Gizis and Reid (\cite{gizis97b}) validated that metallicity scale 
with observations of binary stars containing one M-dwarf component
and one warmer star.

\begin{table}
\caption[]{New double-lined spectroscopic binaries}
\label{table:SB2}
\begin{tabular}{llll}
\hline
Name & HD & $\alpha$ (2000) & $\delta$ (2000)\\
\hline
\object{GJ 3409 B} &263175\,B&06:46:07.6&$+$32:33:13.2\\
\object{Gl 771 B} &- &19:55:18.8&$+$06:24:36\\
\hline
\end{tabular}
\end{table}

\begin{table*}
\centering
\caption[]{Stellar parameters measured on the primaries. [Fe/H] applies for both components.}
\label{table:parameters}
\begin{tabular}{lllllll}
\hline
\multicolumn{2}{c}{Primary} & Secondary &\multicolumn{4}{c}{Stellar parameters measured on the primaries}\\
Gliese/NLTT & HD   &        &\multicolumn{1}{c}{ T$_{eff}$ }      &\multicolumn{1}{c}{ log($g$) }        &\multicolumn{1}{c}{ v$_t$ }            &\multicolumn{1}{c}{[Fe/H]}  \\
\hline
GJ\,1021        &4391         & {\bf CD-48 176B}	& 5967 $\pm$ 70	& 4.74 $\pm$ 0.14 & 1.39 $\pm$ 0.17 & {\bf $-$0.08}   $\pm$  0.09 \\
Gl\,34\,A         &4614         & {\bf Gl\,34\,B}		& 5895 $\pm$ 68	& 4.43 $\pm$ 0.13 & 1.37 $\pm$ 0.22 & {\bf $-$0.31}   $\pm$  0.09 \\
Gl\,53.1A        &6660         & {\bf Gl\,53.1\,B}	& 4705 $\pm$ 131	& 4.33 $\pm$ 0.26 & 0.76 $\pm$ 0.25 & {\bf $+$0.07}  $\pm$  0.12 \\
Gl\,81.1\,A      &11964       & {\bf Gl\,81.1\,B}	& 5311 $\pm$ 42	& 3.97 $\pm$ 0.08 & 0.86 $\pm$ 0.05 & {\bf $+$0.09}  $\pm$  0.06 \\
Gl\,105\,A       &16160       & {\bf Gl\,105\,B}	& 4846 $\pm$ 65	& 4.29 $\pm$ 0.13 & 0.81 $\pm$ 0.10 & {\bf $-$0.19}   $\pm$  0.07 \\
Gl\,107\,A       &16895       & {\bf Gl\,107\,B}	& 6328 $\pm$ 86	& 4.43 $\pm$ 0.17 & 1.73 $\pm$ 0.31 & {\bf $-$0.03}   $\pm$  0.09 \\
GJ\,3194\,A    &18757       & {\bf GJ\,3195\,B}		& 5681 $\pm$ 34	& 4.49 $\pm$ 0.06 & 1.01 $\pm$ 0.07 & {\bf  $-$0.31}  $\pm$  0.04 \\
Gl\,166\,A       &26965       & {\bf Gl\,166\,C}	& 5125 $\pm$ 56	& 4.43 $\pm$ 0.11 & 0.30 $\pm$ 0.15 & {\bf $-$0.33}   $\pm$  0.06 \\
Gl\,211            &37394       & {\bf Gl\,212}		& 5293 $\pm$ 109	& 4.50 $\pm$ 0.21 & 0.79 $\pm$ 0.17 & {\bf $+$0.04}  $\pm$  0.11 \\
Gl\,231.1\,A    &43587       & {\bf Gl\,231.1\,B}	& 5946 $\pm$ 32	& 4.38 $\pm$ 0.06 & 1.15 $\pm$ 0.06 & {\bf $-$0.02}   $\pm$  0.04 \\
Gl\,250\,A       &50281       & {\bf Gl\,250\,B}	& 4670 $\pm$ 80	& 4.41 $\pm$ 0.16 & 0.70 $\pm$ 0.19 & {\bf $-$0.15}   $\pm$  0.09 \\
Gl\,297.2\,A    &68146       & {\bf Gl\,297.2\,B}	& 6280 $\pm$ 106 	& 4.46 $\pm$ 0.21 & 1.81 $\pm$ 0.26 & {\bf $-$0.09}  $\pm$  0.09 \\
Gl\,324\,A       &75732       & {\bf Gl\,324\,B}	& 5283 $\pm$ 59	& 4.36 $\pm$ 0.11 & 0.87 $\pm$ 0.08 & {\bf $+$0.32}  $\pm$  0.07 \\
Gl\,505\,A       &115404     & {\bf Gl\,505\,B}	& 4983 $\pm$ 48	& 4.41 $\pm$ 0.09 & 0.84 $\pm$ 0.07 & {\bf $-$0.25}   $\pm$  0.05 \\
Gl\,544\,A       &125455	& {\bf Gl\,544\,B}	& 5271 $\pm$ 189	& 4.85 $\pm$ 0.37 & 0.87 $\pm$ 0.36 & {\bf $-$0.20}   $\pm$ 0.19 \\
NLTT\,45789 &164595     & {\bf NLTT\,45791}	& 5696 $\pm$ 41	&4.36  $\pm$ 0.08 & 0.83 $\pm$ 0.06 & {\bf $-$0.07}   $\pm$  0.05 \\
Gl\,768.1\,A    &187691     & {\bf Gl\,768.1\,B}	& 6248 $\pm$ 93	& 4.63 $\pm$ 0.18 & 2.36 $\pm$ 0.48 & {\bf $+$0.07}  $\pm$ 0.12 \\
Gl\,783.2\,A    &191785     & {\bf Gl\,783.2\,B}	& 5094 $\pm$ 66	& 4.31 $\pm$ 0.13 & 0.30 $\pm$ 0.19 & {\bf $-$0.16}   $\pm$ 0.08 \\
Gl\,797\,A       &197076\,A & {\bf Gl\,797\,B}	& 5889 $\pm$ 32	& 4.59 $\pm$ 0.06 & 1.01 $\pm$ 0.06 & {\bf $-$0.07}   $\pm$  0.04 \\
Gl\,806.1\,A    &197989     & {\bf Gl\,806.1\,B}	& 4911 $\pm$ 85	& 2.98 $\pm$ 0.17 & 1.61 $\pm$ 0.08 & {\bf $-$0.05}   $\pm$ 0.13 \\
Gl\,872\,A       &215648     & {\bf Gl\,872\,B}	& 6156 $\pm$ 99	& 4.09 $\pm$ 0.19 & 4.05 $\pm$ 2.01 & {\bf $-$0.36}   $\pm$  0.11 \\
\hline
  \end{tabular}
\end{table*}

Valenti et al. (\cite{valenti98}) performed detailed spectral synthesis 
of a very high resolution spectrum of Gl\,725\,B (vB~10) to determine its 
atmospheric parameters. Zboril \& Byrne (\cite{zboril98}) matched 
high resolution red spectra (5500-9000 \AA) of 7~K and 11~M dwarfs 
to Allard \& Hauschildt (\cite{allard95}) synthetic spectra. They conclude 
that for the M~dwarfs the resulting metallicities are only indicative, a 
conclusion that probably applies to most previous references, at least 
for the later subtypes. Jones et al. (\cite{jones02}) synthesized 
the water vapor bands for Infrared Space Observatory (ISO)  2.5-3 $\mu$m
spectra of 3 M dwarfs to derive their parameters.

The limited overlap between these studies shows that they have not yet
converged to consistency. Gl\,725\,B was measured by both Valenti et al. 
(\cite{valenti98}) and Zboril \& Byrne (\cite{zboril98}), who respectively
derive [M/H]=$-$0.92 and [M/H]=$-$0.15. Jones et al. (\cite{jones96}, 
\cite{jones02}), Zboril \& Byrne (\cite{zboril98}) and Dawson \& De 
Robertis (\cite{dawson04}) 
all measured  Barnard's star and found sub-solar to solar metallicity, but
the values spread from $-$0.75 to 0.0 dex. Kapteyn's star is also 
consistently found to be sub-metallic (Mould \cite{mould76}; Jones \cite{jones02};
Woolf \& Wallerstein \cite{woolf04}, \cite{woolf05}), as expected from
its population~II kinematics but again with some dispersion. 

The above references attempt to simultaneously determine the effective 
temperature (T$_{eff}$), the gravity (log~$g$) and the metallicity ([M/H]),
by minimizing the difference between observed and model spectra. 
Unfortunately the 3 parameters are strongly coupled, in particular 
for spectra that are not flux calibrated. Furthermore, the models do 
not yet reproduce the observed spectra in perfect detail, mostly due to 
remaining shortcomings in their molecular transition databases, especially for the later subtypes. The interpretation therefore involves 
estimations as to which features of the spectra should be ignored and 
which should be given maximum weight, and to some extent the process remains 
an art. Different practitioners would likely obtain somewhat different 
answers from the same data and atmospheric models, and they definitely 
do when they analyse different spectral bands observed with different spectral 
resolutions.

The recent analyses of Kapteyn's and Barnard's stars by Woolf \& 
Wallerstein (\cite{woolf04}) and Dawson \& De Robertis (\cite{dawson04}), 
by contrast, are anchored in model-independent T$_{eff}$ and log~$g$ values from
S\'egransan et al. (\cite{segransan03}). S\'egransan et al. 
(\cite{segransan03}) combined their interferometric radius measurements 
with the bolometric flux to determine T$_{eff}$, reversing the more usual 
procedure of determining stellar radii from effective temperature and
luminosity, and they computed the gravity from the linear radius and a mass
derived from the well constrained K~band mass-luminosity relation.
Woolf \& Wallerstein (\cite{woolf04}) and Dawson \& De Robertis 
(\cite{dawson04}) could therefore concentrate the full information 
content of their spectra on determining the metallicity, free of
any uncontrolled coupling with the other atmospheric parameters.

Woolf \& Wallerstein (\cite{woolf05}, hereafter WW05) analysed a much
larger sample of 35 K and M dwarfs, that for now do not have 
interferometric radius measurements. WW05 therefore rely on 
photometric effective temperatures (T$_{eff}$) and they use a photometric 
radius to compute the gravity. While less direct than the S\'egransan et 
al. (\cite{segransan03}) measurements, this procedure rests on relations 
which that paper validates, and that in our view is currently preferable
to determining those parameters from the spectrum. The 15 Woolf \& 
Wallerstein (\cite{woolf05}) M dwarfs are overwhelmingly of early subtypes 
(only one is later than M1.5) and they concentrate on low metallicity
targets. Their spectra therefore have limited molecular veiling. Together 
with their use of the latest generation of the PHOENIX models, this
reduces their sensitivity to the remaining shortcomings of the molecular 
opacity databases. The parameter space which they cover complements our own 
measurements, and we make extensive use of these data in our discussion.


\section{Observations and analysis}
\label{sect:data analysis}

\subsection{Sample, observation and data reduction}
We selected wide physical visual binaries composed of an F-, G- or K-primary 
component and an M-dwarf secondary, from the Gliese \& Jareiss 
(\cite{gliese91}) catalogue of nearby stars, the Poveda et al. 
(\cite{poveda94}) catalogue of wide-binary and multiple systems
 of nearby stars, and the Gould \& Chanam\'e (\cite{gould04}) list
of physical HIPPARCOS binaries. We further required that the components
be separated by at least $5''$ and that the secondary be brighter than 
V=14, to facilitate observations of the faint M dwarf. Fast rotators, 
double-lined spectroscopic binaries (SB2) and close visual binaries were 
rejected {\it a priori} when known, and otherwise discarded {\it a 
posteriori}. These criteria resulted in a list of 70 pairs. 

We discarded 2 systems whose secondaries (\object{GJ 3409 B} and 
\object{Gl 771 B}) were SB2 binaries (reported in Table 
\ref{table:SB2}). The \object{Gl 549} system had to be rejected
as the F7V primary is a fast rotator 
(v$\sin i \sim$ 50~km.s$^{-1}$). The \object{Gl 695} system was also
rejected as both components are themselves close
visual binaries. Here we analyse 21 of those systems 
(Table \ref{table:targets}),
of which 20 have M-dwarf secondaries (the last one being classified
as K7V). 

Most of the spectra were gathered using the ELODIE 
spectrograph (Baranne et al. \cite{baranne96}) on the 1.93-m telescope
of Observatoire de Haute-Provence (France). ELODIE covers 
a visible spectral range from 3850 to 6800\,\AA~with a resolution of 45000. 
For \object{GJ 1021}, \object{Gl 166 A} and \object{Gl 250 A} we reuse
spectra observed by Santos et al. (\cite{santos01}) with the CORALIE 
spectrograph (1.20-m Swiss Telescope, La Silla Observatory ESO, Chile). 
CORALIE has a slightly wider spectral range than ELODIE, 3650 to 6900\,\AA, 
and a slightly higher resolution of 50000. On-line processing is integrated 
with control software of both spectrographs, and automatically produces 
optimally extracted, flat-fielded and wavelength calibrated spectra, with 
algorithms described in Baranne et al. (\cite{baranne96}). For all primaries 
the present observations used the ``Object-only'' mode of the spectrograph, 
where its optional reference fiber is not illuminated. This mode provides 
optimal scattered light correction, at the cost of degraded radial velocity 
precision ($\sim$100~m.s$^{-1}$). The wavelength calibration used a single 
Thorium-Argon exposure obtained at the beginning of each night.
For each primary we recorded a sequence of 3 spectra, and applied 
a median filter to remove any unflagged cosmic ray hit. The combined 
spectra have signal-to-noise ratios of approximately 200 per pixel 
($\sim$300 per resolution element), amply sufficient for our 
spectroscopic analysis. We also obtained spectra for the secondaries, 
usually with a much lower signal to noise ratio, from which we planned 
to derive spectroscopic metallicity diagnostics that can be applied at
moderate/low signal to noise ratio data. That goal has proved more difficult
than we expected, and it will be discussed in a future paper if we are
successful.

\begin{figure}
        \centering
        \includegraphics[width=0.45\textwidth]{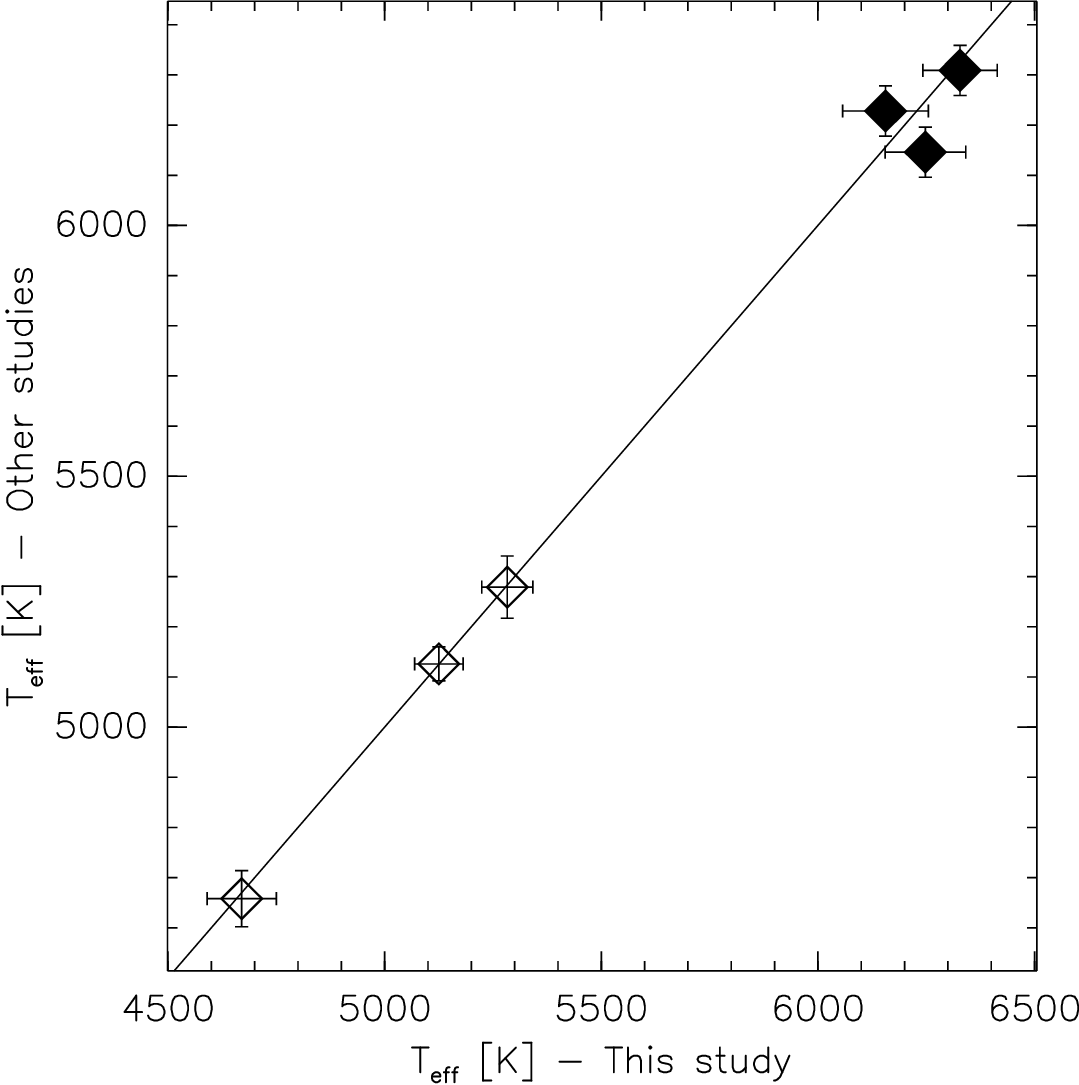}
        \includegraphics[width=0.45\textwidth]{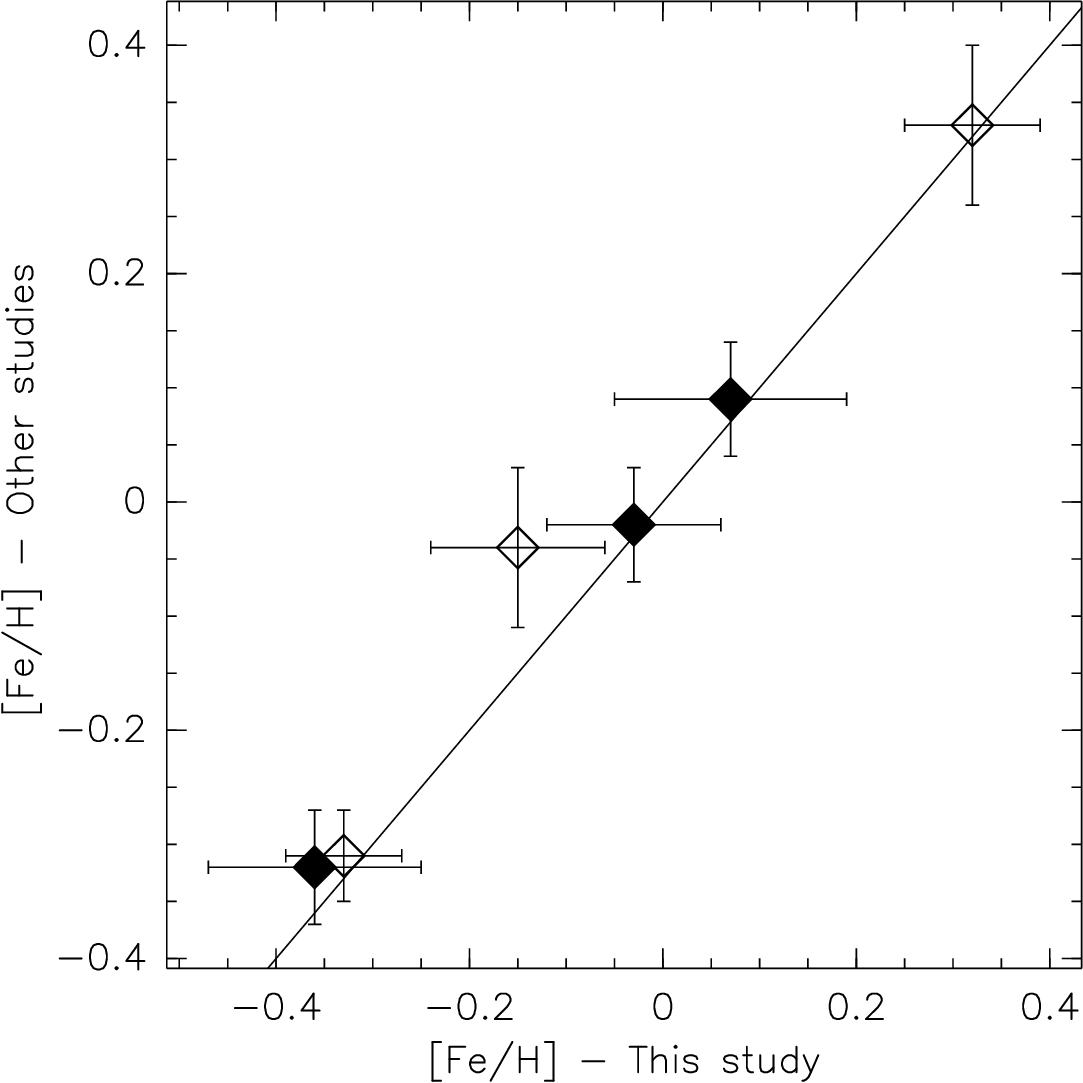}
                \caption{Comparison of T$_{eff}$ (upper panel) and 
[Fe/H] (lower panel) between our study and Santos et al. (\cite{santos04}, 
open diamonds) and Edvardsson et al. (\cite{edvardsson93}, filled diamonds). 
The errorbars represent the individual 1~$\sigma$ errors for our study and
for Santos et al. Edvardsson et al. do not list individual error estimates and
we adopt their typical errors of 50 K for T$_{eff}$ and 0.05 dex for 
[Fe/H]. The over-plotted lines represent an identity relation, not
a fit to the data.}
        \label{table:comparison}
\end{figure}

\begin{figure*}[ht]
        \centering
	\includegraphics[width=0.45\textwidth]{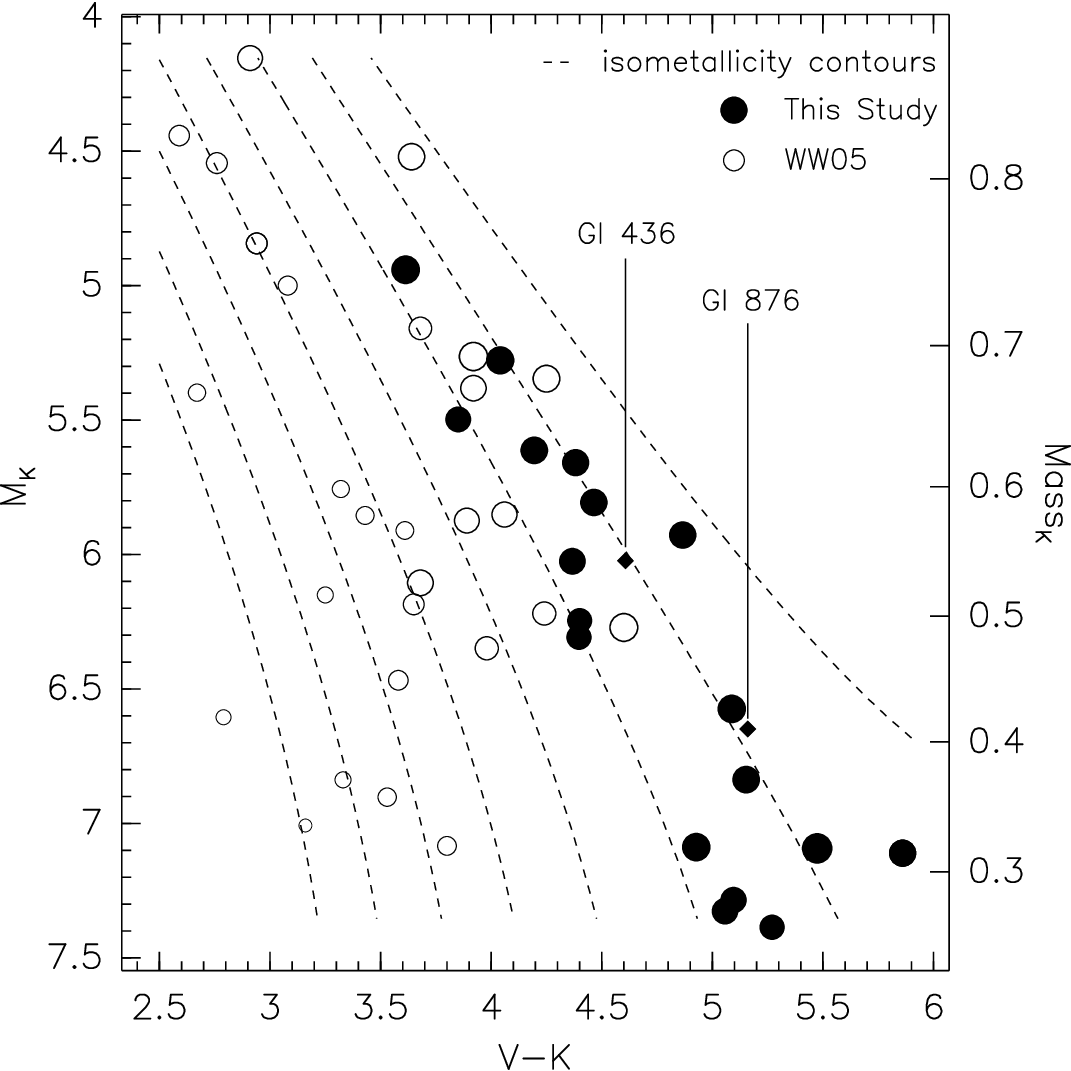}
        \hspace{0.08\textwidth}
	\includegraphics[width=0.45\textwidth]{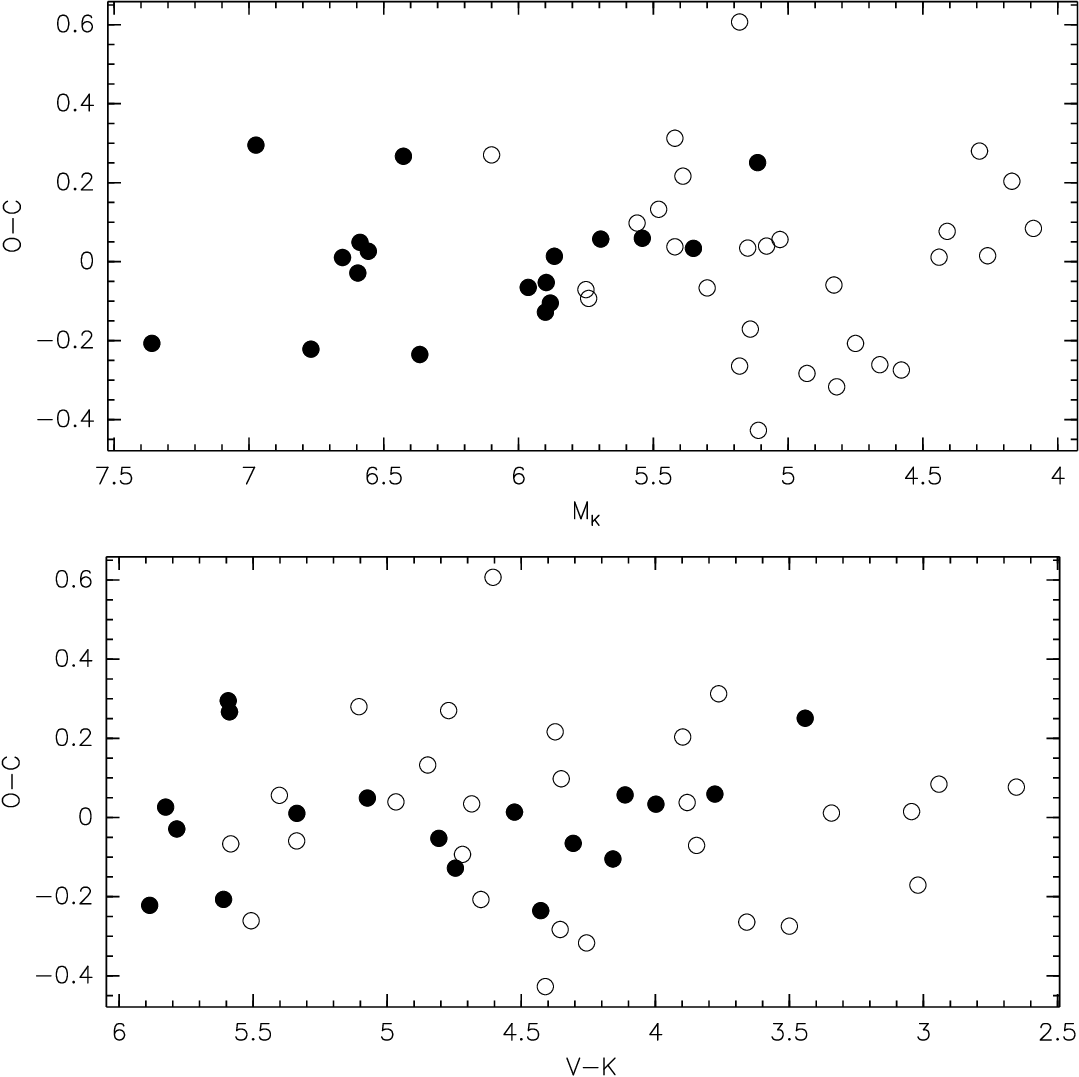}
	\caption[]{{\it Left panel:} Color-magnitude diagram V$-$K
          vs. M$_K$.  The filled circles correspond to our metallicity
          determinations and the open circles to those from WW05. The
          symbol size is proportional to the metallicity. The dashed lines
          represent isometallicity contours for the polynomial
            relation of Eq. \ref{eq:calib}, spaced by 0.25~dex from
          $-$1.50~dex (left) to $+$0.25~dex (right). The right-hand axis
          shows masses from the DFS00 K-band Mass-luminosity, which has
          very low dispersion and allows to interpret the figure as a
          Mass-Colour-Metallicity diagram. Gl\,876 and Gl\,436, the two
          known M-dwarf planet-host stars, are indicated to illustrate
          their solar metallicity.  {\it Right panels:} Residuals from the
          calibration as a function of both M$_K$ and V$-$K photometry.}
	\label{fig:vkk}
\end{figure*}

\subsection{Spectroscopic analysis}
Our spectroscopic analysis of the primaries follows the procedure 
described by Santos et al. (\cite{santos04}) for planet host stars.
Briefly, we used the Gaussian fitting procedure of the IRAF 
{\verb splot } task to the measure equivalent widths (${W}_\lambda$) 
of 39 \ion{Fe}{I} and 12 \ion{Fe}{II} lines. The stellar parameters 
were then derived using the 2002 version of the MOOG code (Sneden 
\cite{sneden73}), and a grid of Kurucz (\cite{kurucz93}) ATLAS9 model 
atmospheres. Table~\ref{table:parameters} presents the resulting
atmospheric parameters (microturbulence, effective temperature (T$_{eff}$), 
surface gravity and iron abundance ([Fe/H]), which we use as a measurement 
of the overall metallicity ([M/H]).

The standard errors on T$_{\mathrm{eff}}$, $\log{g}$, $\xi_t$ and 
[Fe/H] were derived as described in Santos et al. (\cite{santos04}), 
following the prescriptions of Gonzalez \& Vanture (\cite{gonzalez98}). 
The resulting uncertainties are internal, in the sense that they ignore 
possible scale offsets. There is currently some disagreement on e.g. 
the apropriate temperature scale for solar-type dwarfs, as well as which model atmospheres
better reproduce the real stellar atmospheres. The true errors may
consequently be larger, but the listed standard errors are appropriate
for comparisons within our sample. As discussed in Santos 
et al. (\cite{santos04}, \cite{santos05}), the method and the grid of ATLAS9 atmospheres used 
gives excellent results, compatible with those derived by
other authors using other model atmospheres and methods to derive the
stellar parameters and metallicities.

Six of the observed primaries have published stellar parameters (Santos et al. 
\cite{santos04}; Edvardsson et al. \cite{edvardsson93}). Comparison of our 
determinations of T$_{eff}$ and [Fe/H] with these litterature values 
(Fig.~\ref{table:comparison}) shows that they agree to within the stated 
errors.

\section{A photometric calibration of M-dwarf metallicities}
\label{sect:calibration}
From this point on, we use a sample (Table~4) that combines our own 
metallicity measurements with those of WW05. As discussed above, we 
expect the latter to be reliable, and they reach to lower metallicities 
for mostly hotter atmospheres. The two datasets are therefore complementary,
but they have enough overlap in the (T$_{eff}$,[Fe/H]) plane to assess 
possible systematic differences (Fig.~\ref{fig:vkk}). Table~4 contains all 
WW05 stars with known parallaxes, except \object{LHS 1138}, listed as a 
G5 dwarf in SIMBAD, and \object{GJ 1064 D}, a clear outlier in our 
relations and 
perhaps a photometric binary. We restrict the analysis to stars brighter 
than M$_K$=7.5, since the sampling is very sparse for fainter stars. 

The left panel of Fig.~\ref{fig:vkk} displays the effect of metallicity 
in the M$_K$ vs V$-$K observational Hertzsprung-Russell diagram, with symbol 
sizes proportional 
to the metallicity of the corresponding stars. After experimenting with 
several colour-magnitude diagrams, we found that amongst commonly available 
photometric bands this combination maximizes the metallicity sentivity.
It is immediately obvious that lower metallicity stars are much bluer
at a given absolute M$_K$ magnitude, and we find the metallicity well 
described by the following polynomial relation between M$_K$ and V$-$K:
\begin{eqnarray}
[Fe/H] &= &0.196 -1.527\,M_K+0.091\,M_K^2\nonumber\\
&&+1.886(V-K) -0.142(V-K)^2,
\label{eq:calib}
\end{eqnarray}
valid for $M_K \in [4,7.5]$, $V-K\in [2.5,6]$ and $[Fe/H]\in [-1.5,+0.2]$
and with an observed dispersion of only 0.2~dex.

Part of this dispersion might be due to a few of the higher mass 
stars having evolved slightly off the main sequence. For instance between 
8~Gyr and 10~Gyr  an 0.8~M$_\odot$ star brightens by $\sim$0.3~mag in the 
V band and $\sim$0.2~mag in the K band, moving noticeably in the 
Fig.~\ref{fig:vkk} diagram. By 0.7~M$_\odot$ stellar evolution effects
become small, with a brightening between 8~Gyr to 12~Gyr of $\sim$0.1~mag 
in both the V and K bands. The age/metallicity relation might therefore
introduce a small systematic bias in our relation, but that would affect
at most the highest mass fringe of its validity range.

The lower panels of Fig.~\ref{fig:vkk} display the residuals from that
relation. 
The absence of any obvious systematic pattern demonstrates that 
the calibration remains valid over its stated range. The consistency 
between the residuals of the WW05 measurements and ours ensures 
that any systematic difference between the two datasets must be 
small where they overlap, for approximately solar metallicities.
For significantly subsolar metallicities (i.e. well below -0.25~dex)
we have no independent validation of the WW05 data. It should
be noted however that their approach has maximal uncertainties for high 
metallicities, where molecular veiling is most severe. The good agreement
where difficulties would be most expected suggests that the low metallicity
data points are valid as well.


\section{The V-band mass-luminosity relation}
\label{sect:mlfunction}

\begin{figure}
	\centering
	\includegraphics[width=0.45\textwidth]{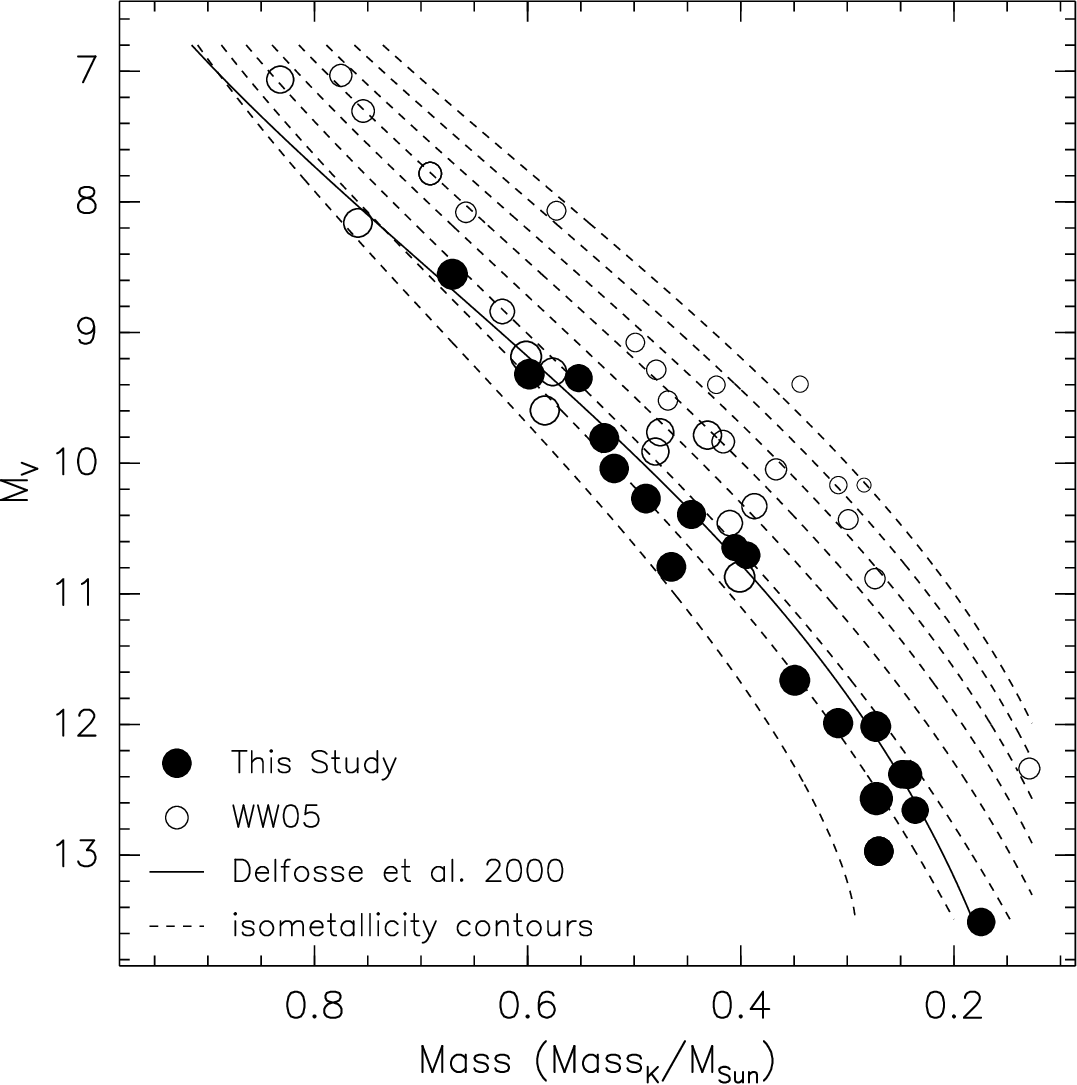}
	\caption[]{V band M/L relation, with masses derived from the K-band 
M/L relation of DFS00 and 2MASS photometry. The filled circles 
represent our metallicity determinations and the open circles those from
WW05. The symbol size is proportional to the metallicity, and the dashed 
contours represent isometallicity for the Eq. \ref{eq:calib}
calibration, spaced by 0.25~dex from +0.25 (left) to -1.5~dex (right). 
The solid lines represents the V-band empirical M/L relation of DFS00.}
	\label{fig:VMk}
\end{figure}

\begin{figure}
	\centering
	\includegraphics[width=0.45\textwidth]{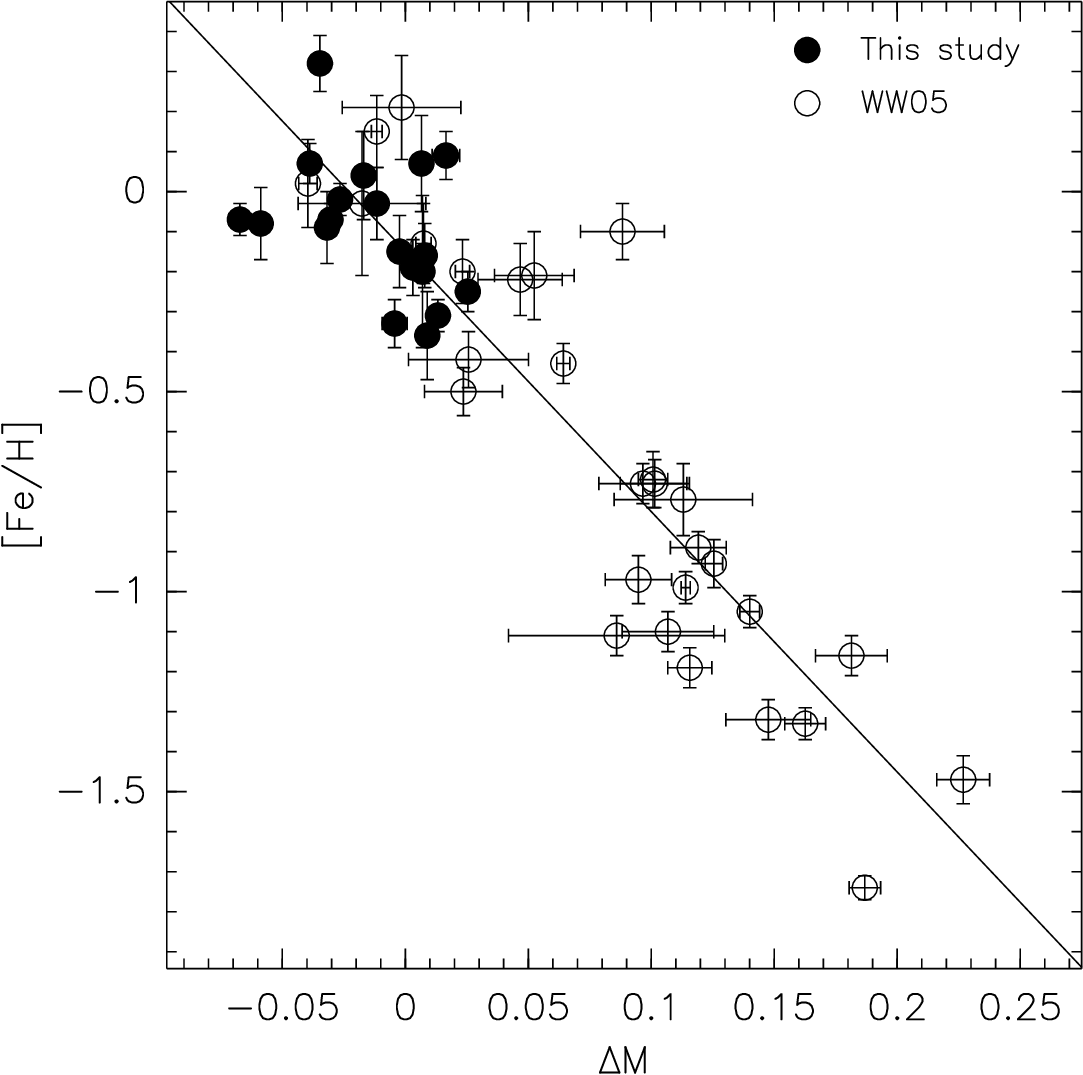}
	\caption[]{Metallicity of M and K dwarfs (filled circles for
our measurements, and open circles for WW05 data) as a function of 
the difference ($\Delta$M) between masses calculated from the V- and K-band 
M/L relations of DFS00.}
	\label{fig:fehmvmk}
\end{figure}

\begin{table*}[hp]
\begin{center}
\caption[]{Apparent magnitudes, parallaxes, masses derived from the M/L
relations of DFS00, and metallicities from this study and from WW05.}
\label{table:db}
\begin{tabular}{llr@{\,$\pm$\,}rcr@{\,$\pm$\,}rr@{\,$\pm$\,}lcccr@{\,$\pm$\,}rc} 
      \hline
Star & Spectral&\multicolumn{2}{c}{V} & source &\multicolumn{2}{c}{K} & \multicolumn{2}{c}{$\pi$} & source& Mass$_{V}$ & Mass$_{K}$&\multicolumn{2}{c}{[Fe/H]} & source\\
        & type&\multicolumn{2}{c}{[mag.]} &V$^\dagger$ &\multicolumn{2}{c}{[mag.]} & \multicolumn{2}{c}{[mas]} &$\pi^{\dagger\dagger}$ &[M$_{\odot}$] & [M$_{\odot}$] & \multicolumn{2}{c}{[dex]}&[Fe/H]$^\ddagger$\\
\hline
CD-48\,176B     &M              &\multicolumn{2}{c}{13.50} &S &7.64   &0.02   &66.92  &0.73   &H      &0.212  &0.270  &$-$0.08        &0.09   &a\\ 
GJ\,3195\,B     &M3             &\multicolumn{2}{c}{12.50} &G &8.10   &0.03   &43.74  &0.84   &H      &0.408  &0.395  &$-$0.31        &0.04   &a \\
GJ\,3825        &esdM1.5        &14.55  &0.03   &M &10.86  &0.01   &36.1   &3.2    &Y      &0.255  &0.129  &$-$0.93        &0.06   &b \\
GJ\,687         &M3             &9.15   &0.03   &M &4.55   &0.02   &220.9  &0.9    &H      &0.389  &0.401  &$+$0.15        &0.09   &b\\ 
GJ\,9192        &K4             &10.70  &0.02   &M &7.76   &0.02   &26.1   &2.1    &H      &0.793  &0.691  &$-$0.73        &0.06   &b \\
GJ\,9371        &sdM0.0         &12.20  &0.03   &M &8.67   &0.02   &44.3   &2.8    &H      &0.439  &0.299  &$-$1.05        &0.04   &b \\
G\,22-15        &K5V            &9.23   &0.02   &M &6.47   &0.02   &41.2   &1.3    &H      &0.855  &0.754  &$-$0.72        &0.07   &b \\
Gl\,105\,B      &M4             &11.67  &0.01   &M &6.57   &0.02   &138.72 &1.04   &H      &0.252  &0.248  &$-$0.19        &0.07   &a \\
Gl\,107\,B      &M1.5           &\multicolumn{2}{c}{10.06}  &G &5.87   &0.02   &89.03  &0.79   &H      &0.517  &0.528  &$-$0.03        &0.09   &a \\
Gl\,166\,C      &M4.5           &\multicolumn{2}{c}{11.17}  &G &5.9    &0.10   &198.25 &0.84   &H      &0.232  &0.236  &$-$0.33        &0.06   &a \\
Gl\,191         &sdM1.0         &8.85   &0.03   &M &5.05   &0.02   &255.1  &0.9    &H      &0.388  &0.274  &$-$0.99        &0.04   &b \\
Gl\,205         &M1.5           &7.96   &0.01   &M &4.04   &0.26   &175.7  &1.2    &H      &0.600  &0.601  &$+$0.21        &0.13   &b \\
Gl\,212         &M0.5           &9.80   &0.01   &T &5.76   &0.02   &80.13  &1.67   &H      &0.572  &0.598  &$+$0.04        &0.11   &a \\
Gl\,231.1\,B    &M3.5           &\multicolumn{2}{c}{13.42}  &G &8.28   &0.02   &51.76  &0.78   &H      &0.282  &0.309  &$-$0.02        &0.04   &a \\
Gl\,250\,B      &M2             &\multicolumn{2}{c}{10.09}  &G &5.72   &0.04   &114.94 &0.86   &H      &0.442  &0.446  &$-$0.15        &0.09   &a \\
Gl\,297.2\,B    &M2             &11.80  &0.01   &M &7.42   &0.02   &44.47  &0.77   &H      &0.484  &0.519  &$-$0.09        &0.09   &a \\
Gl\,324\,B      &M3.5           &\multicolumn{2}{c}{13.14} &G &7.67   &0.02   &76.8   &0.84   &H      &0.239  &0.273  &$+$0.32        &0.07   &a \\
Gl\,380         &K5             &6.60   &0.02   &M &2.96   &0.29   &205.2  &0.8    &H      &0.742  &0.759  &$-$0.03        &0.18   &b \\
Gl\,411         &M2V            &7.49   &0.02   &M &3.25   &0.31   &392.5  &0.9    &H      &0.436  &0.410  &$-$0.42        &0.07   &b\\ 
Gl\,412\,A      &M0.5           &8.75   &0.04   &M &4.77   &0.02   &206.9  &1.2    &H      &0.451  &0.387  &$-$0.43        &0.05   &b \\
Gl\,414\,B      &M1.5           &9.98   &0.04   &M &5.73   &0.02   &83.8   &1.1    &H      &0.544  &0.584  &$+$0.02        &0.11   &b\\ 
Gl\,505\,B      &M0.5           &\multicolumn{2}{c}{9.60}   &G &5.75   &0.02   &89.07  &0.99   &H      &0.554  &0.552  &$-$0.25        &0.05   &a \\
Gl\,506.1       &sdK            &10.84  &0.02   &M &8.17   &0.02   &27.9   &2.5    &H      &0.754  &0.573  &$-$1.16        &0.05   &b\\ 
Gl\,526         &M1.5           &8.46   &0.01   &M &4.42   &0.21   &184.1  &1.3    &H      &0.520  &0.431  &$-$0.10        &0.07   &b \\
Gl\,53.1\,B     &M3             &\multicolumn{2}{c}{13.60} &G &8.67   &0.02   &48.2   &1.06   &H      &0.280  &0.273  &$+$0.07        &0.12   &a\\ 
Gl\,544\,B      &M6             &\multicolumn{2}{c}{15.10}  &G &9.59   &0.02   &48.12  &1.11   &H      &0.215  &0.174  &$-$0.20        &0.19   &a \\
Gl\,701         &M1             &9.37   &0.03   &M &5.31   &0.02   &128.3  &1.4    &H      &0.503  &0.480  &$-$0.20        &0.08   &b\\ 
Gl\,768.1\,B    &M3.5           &\multicolumn{2}{c}{13.10} &G &8.01   &0.03   &51.57  &0.77   &H      &0.310  &0.349  &$+$0.07        &0.12   &a \\
Gl\,783.2\,B    &M4             &\multicolumn{2}{c}{13.94} &G &8.88   &0.02   &48.83  &0.91   &H      &0.251  &0.243  &$-$0.16        &0.08   &a \\
Gl\,797\,B      &M2.5           &\multicolumn{2}{c}{11.88} &G &7.42   &0.02   &47.65  &0.76   &H      &0.460  &0.489  &$-$0.07        &0.04   &a \\
Gl\,809         &M0.5           &8.54   &0.04   &M &4.62   &0.02   &142.0  &0.8    &H      &0.584  &0.576  &$-$0.13        &0.10   &b \\
Gl\,81.1\,B     &K7             &\multicolumn{2}{c}{11.21} &G &7.60   &0.03   &29.43  &0.91   &H      &0.684  &0.671  &$+$0.09        &0.06   &a\\ 
Gl\,872\,B      &M1             &\multicolumn{2}{c}{11.70} &G &7.30   &0.02   &61.54  &0.77   &H      &0.414  &0.405  &$-$0.36        &0.11   &a \\
Gl\,887         &M0.5           &7.35   &0.02   &M  &3.46   &0.20   &303.9  &0.9    &H      &0.522  &0.476  &$-$0.22        &0.09   &b \\
LHS\,104        &esdK7          &13.74  &0.02   &M  &10.4   &0.02   &19.3   &3.0    &Y      &0.471  &0.308  &$-$1.33        &0.04   &b \\
LHS\,12         &M0.5           &12.26  &0.04   &M  &8.68   &0.02   &36.1   &4.3    &H      &0.486  &0.367  &$-$0.89        &0.04   &b \\
LHS\,170        &sdK            &10.68  &0.01   &M  &7.60   &0.02   &30.2   &2.4    &H      &0.753  &0.658  &$-$0.97        &0.06   &b \\
LHS\,173        &sdK7           &11.11  &0.01   &M  &7.79   &0.02   &39.2   &2.5    &H      &0.615  &0.499  &$-$1.19        &0.05   &b \\
LHS\,174        &sdM0.5         &12.75  &0.01   &M  &9.14   &0.02   &22.6   &7.4    &Y      &0.554  &0.468  &$-$1.11        &0.05   &b \\
LHS\,1819       &K4             &10.88  &0.02   &M  &8.29   &0.03   &17.0   &2.6    &H      &0.888  &0.775  &$-$0.77        &0.09   &b \\
LHS\,1841       &K              &13.18  &0.03   &M  &10.39  &0.02   &17.5   &3.3    &Y      &0.571  &0.344  &$-$1.47        &0.06   &b\\ 
LHS\,236        &sdK7           &13.10  &0.01   &M  &9.85   &0.02   &18.2   &2.9    &Y      &0.570  &0.423  &$-$1.32        &0.05   &b \\
LHS\,2938       &K7             &10.67  &0.02   &M  &7.76   &0.02   &19.0   &2.0    &H      &0.885  &0.832  &$-$0.21        &0.11   &b\\ 
LHS\,3084       &sdK            &13.43  &0.03   &M  &9.78   &0.02   &19.1   &2.9    &Y      &0.513  &0.416  &$-$0.73        &0.05   &b\\ 
LHS\,343        &sdK            &13.82  &0.02   &M  &10.66  &0.02   &18.6   &3.7    &Y      &0.471  &0.284  &$-$1.74        &0.03   &b \\
LHS\,467        &esdK7          &12.21  &0.03   &M  &8.78   &0.02   &26.0   &3.6    &H      &0.586  &0.479  &$-$1.10        &0.05   &b \\
LHS\,5337       &M0             &11.15  &0.03   &M  &7.47   &0.02   &34.5   &3.3    &H      &0.647  &0.624  &$-$0.50        &0.06   &b\\ 
NLTT\,45791     &M              &\multicolumn{2}{c}{13.10} &N &8.23   &0.03   &34.57  &0.73   &G      &0.398  &0.465  &$-$0.07        &0.05   &a \\
\hline
  \end{tabular}
\end{center}
$^\dagger$ M: Mermilliod, Mermilliod \& Hauck (\cite{mermilliod97}), G: Gliese \& Jareiss (\cite{gliese91}), N: Gould \& Chanam\'e (\cite{gould04}), T: Tycho Input Catalogue (Egret et al. \cite{egret92}), S: Simbad database. Unfortunately their is no uncertainties on V-band photometry in the Gliese \& Jareiss catalogue.\\
$^{\dagger\dagger}$ H: ESA (\cite{esa97}), Y: van Altena, Lee, \& Hoffleit 
(\cite{vanaltena95}), G: Gould \& Chanam\'e (\cite{gould04})\\
$^\ddagger$ a: this paper, b: Woolf \& Wallerstein (\cite{woolf05})\\
The K-band photometry is from 2MASS. The mass estimates use the listed photometry and the M/L relations of Delfosse et al. (\cite{delfosse00})\\
\end{table*}

As demonstrated by DFS00, the infrared J-, H- and K-band M/L relations are
very tight and in excellent agreement with model predictions, while the
V-band relation has a large intrinsic scatter. The contrasting dispersions
were qualitatively expected from different metallicity sensitivities for
the visual and infrared bands (e.g. Chabrier \& Baraffe \cite{chabrier00}),
but the extent of the effect was a surprise to most observers. Metallicity
affects luminosity through a given photometric filter in two ways. 
First, higher metallicity decreases the bolometric luminosity for a given mass,
and second, it shifts flux from the visible range to the near-IR through
higher line-blanketing by \element{TiO} and \element{VO} molecular
bands. The two mechanisms work together to decrease the luminosity 
of the more metal-rich stars through visible filters. In the near-IR by
contrast, the redward shift of the flux distribution of the metal-rich
stars counteracts their lower bolometric luminosity. The models therefore
predict IR absolute magnitudes that are largely insensitive to metallicity,
and the tight empirical M/L relations confirm this. DFS00 could on the
other hand not quantitatively verify their suggestion that metallicity
explains the V-band dispersion.  The Table~\ref{table:db} measurements now
allow us to perform this verification.

Since the K-band M/L relation is so tight, we can use the parallaxes and 2MASS 
photometry to derive accurate masses. Fig.~\ref{fig:VMk} displays those 
masses (Mass$_{K}$) as a function of the M$_V$ absolute magnitude, with 
symbol sizes proportional to the measured metallicity. 
The figure also shows the DFS00 V-band M/L relation, and isometallicity 
contours obtained by remapping Eq. \ref{eq:calib} to the Mass/M$_V$ plane. 
It is immediately obvious that the position relative to the average M/L
relation correlates with metallicity, with the smallest symbols far 
above the M/L relation and the largest ones under that relation. 

Fig.~\ref{fig:fehmvmk} provides a more quantitative view, by projecting
the Mass/M$_V$/[Fe/H] information along the average V-band M/L relation. 
This diagram of [Fe/H] as a function of the difference between masses
derived from the V- and K-band M/L relations shows a well-defined linear 
correlation 
($[Fe/H] = -0.149 -6.508\,\Delta M,~\sigma([Fe/H])=0.21$). 
This demonstrates  i) that the observed dispersion indeed results
primarily from a metallicity effect, and ii) that the luminosity shift 
for a given metallicity is, to first order, constant between 0.8 and 
0.2 M$_\odot$.

We now have all the elements in hand to examine how the V-band luminosity 
depends on mass and metallicity, and to compute a 
mass-\emph{metallicity}-luminosity relation for very-low-mass stars. We 
find that the V-band luminosity is well described by the following 
polynomial relation:
\begin{eqnarray}
M_V &= &  15.844 -16.534\,Mass  +13.891\,Mass^2\nonumber\\
&&-7.411\,Mass^3+1.153\,[Fe/H]
\label{eq:mml}
\end{eqnarray}
for $Mass \in [0.2\,M_\odot,0.8\,M_\odot]$ and $[Fe/H] \in [-1.5, 0.2]$,
and with a dispersion of 0.28 mag. 

\begin{table*}
\begin{center}
\caption[]{Magnitudes, parallaxes, corresponding masse and metallicity estimates of M-dwarf neighbors}
\label{table:feh9pc}
\begin{tabular}{llr@{\,$\pm$\,}lcl@{\,$\pm$\,}rl@{\,$\pm$\,}rlll}
      \hline
Star & Spectral&\multicolumn{2}{c}{V}&source&\multicolumn{2}{c}{K}& \multicolumn{2}{c}{$\pi$}  & M$_{\star,V}$ & M$_{\star,K}$&[Fe/H]\\
        & type        &\multicolumn{2}{c}{[mag.]} &V$^\dagger$&\multicolumn{2}{c}{[mag.]} & \multicolumn{2}{c}{[mas]} &[M$_{\odot}$]   & [M$_{\odot}$] & [dex]\\
\hline
Gl15A           &M1     &8.10   &0.02 &M&4.02   &0.02   &280.30 &1.00   &0.450  &0.404  &$-$0.45\\
Gl48            &M3     &10.05  &0.01 &M&5.45   &0.02   &122.80 &1.20   &0.438  &0.471  &$+$0.04\\
Gl109           &M3     &10.60  &0.01 &M&5.96   &0.02   &132.40 &2.50   &0.354  &0.350  &$-$0.20\\
Gl205           &M1.5   &7.97   &0.01 &M&4.04   &0.26   &175.70 &1.20   &0.594  &0.602  &$-$0.09\\
LHS1805         &M3.5   &\multicolumn{2}{c}{11.71}      &G&6.64   &0.02   &132.10 &4.90   &0.257  &0.253  &$-$0.16\\
Gl251           &M3V    &10.02  &0.01 &M&5.28   &0.02   &181.30 &1.90   &0.346  &0.350  &$-$0.16\\
Gl273           &M3.5   &9.84   &0.01 &M&4.86   &0.02   &263.30 &1.40   &0.289  &0.291  &$-$0.16\\
Gl285           &M4.5   &11.21  &0.02 &M&5.70   &0.02   &168.60 &2.70   &0.262  &0.309  &$+$0.07\\
GJ1105          &M3.5   &12.01  &0.01 &M&6.88   &0.03   &120.80 &4.40   &0.249  &0.248  &$-$0.15\\
GJ2066          &M2     &10.12  &0.01 &M  &5.77   &0.02   &109.20 &1.80   &0.456  &0.459  &$-$0.14\\
Gl382           &M1.5   &9.26   &0.01 &M&5.01   &0.02   &128.00 &1.50   &0.526  &0.541  &$-$0.02\\
Gl388           &M3     &9.41   &0.02 &M&4.59   &0.02   &204.60 &2.80   &0.391  &0.423  &$+$0.05\\
Gl393           &M2     &9.66   &0.01 &M&5.31   &0.02   &138.30 &2.10   &0.454  &0.448  &$-$0.17\\
Gl402           &M4     &11.66  &0.01 &M&6.37   &0.02   &145.90 &3.80   &0.249  &0.260  &$-$0.06\\
Gl408           &M2.5   &10.04  &0.01 &M&\multicolumn{2}{c}{5.34$^\ddagger$}      &151.00 &1.60   &0.387  &0.372  &$-$0.24\\
Gl411           &M2     &7.49   &0.04 &M&3.25   &0.31   &392.40 &0.90   &0.435  &0.410  &$-$0.33\\
Gl412A          &M0.5   &8.82   &0.01 &M&4.77   &0.02   &206.90 &1.20   &0.458  &0.387  &$-$0.51\\
Gl424           &M0     &9.30   &0.01 &M&5.53   &0.02   &109.90 &1.10   &0.556  &0.502  &$-$0.47\\
Gl445           &M3.5   &10.84  &0.01 &M&5.95   &0.03   &185.50 &1.40   &0.270  &0.247  &$-$0.26\\
Gl450           &M1     &9.74   &0.03 &M&5.61   &0.02   &116.90 &1.40   &0.487  &0.462  &$-$0.29\\
Gl486           &M3.5   &11.38  &0.01 &M&6.36   &0.02   &121.80 &2.90   &0.305  &0.316  &$-$0.08\\
Gl514           &M0.5   &9.04   &0.01 &M&5.04   &0.03   &131.10 &1.30   &0.545  &0.526  &$-$0.23\\
LHS2784         &M3.5   &11.97  &0.02 &W&6.98   &0.02   &109.90 &3.20   &0.265  &0.261  &$-$0.19\\
Gl526           &M1.5   &8.48   &0.03 &M&4.42   &0.02   &184.10 &1.30   &0.527  &0.502  &$-$0.24\\
Gl555           &M4     &11.30  &0.02 &M&5.94   &0.03   &163.50 &2.80   &0.252  &0.284  &$+$0.01\\
Gl581           &M3     &10.55  &0.01 &M&5.84   &0.02   &159.50 &2.30   &0.324  &0.307  &$-$0.25\\
Gl625           &M1.5   &10.11  &0.01 &M&5.83   &0.02   &151.90 &1.10   &0.375  &0.323  &$-$0.48\\
Gl628           &M3.5   &10.07  &0.02 &M&5.08   &0.02   &234.50 &1.80   &0.294  &0.296  &$-$0.12\\
Gl687           &M3     &9.16   &0.02 &M&4.55   &0.02   &220.80 &0.90   &0.343  &0.401  &$+$0.11\\
Gl686           &M1     &9.60   &0.02 &M&5.57   &0.02   &123.00 &1.60   &0.486  &0.447  &$-$0.40\\
Gl701           &M1     &9.36   &0.01 &M&5.31   &0.02   &128.30 &1.40   &0.499  &0.481  &$-$0.30\\
GJ1230B         &M5     &\multicolumn{2}{c}{14.00}   &G&6.62   &0.02   &120.90 &7.20   &0.143  &0.281  &$+$0.15\\
Gl725A          &M3     &8.95   &0.01 &M&4.43   &0.02   &280.30 &3.60   &0.361  &0.334  &$-$0.31\\
Gl725B          &M3.5   &9.72   &0.01 &M&5.00   &0.02   &284.50 &5.00   &0.292  &0.250  &$-$0.34\\
Gl745A          &M1.5   &\multicolumn{2}{c}{10.76}   &G&6.52   &0.02   &115.90 &2.50   &0.373  &0.308  &$-$0.54\\
Gl745B          &M2     &\multicolumn{2}{c}{10.75}   &G&6.52   &0.02   &112.80 &2.40   &0.380  &0.317  &$-$0.52\\
Gl752A          &M2.5   &9.13   &0.01 &M&4.67   &0.02   &170.30 &1.40   &0.460  &0.484  &$-$0.02\\
Gl793           &M2.5   &10.63  &0.01 &M&5.93   &0.02   &125.60 &1.10   &0.376  &0.374  &$-$0.12\\
Gl809           &M0.5   &8.55   &0.02 &M&4.62   &0.02   &141.90 &0.80   &0.589  &0.577  &$-$0.16\\
Gl849           &M3.5   &10.37  &0.01 &M&5.59   &0.02   &114.00 &2.10   &0.410  &0.475  &$+$0.14\\
Gl860A          &M3     &\multicolumn{2}{c}{9.85}    &G&4.78   &0.03   &247.50 &1.50   &0.292  &0.322  &$-$0.04\\
Gl873           &M3.5   &10.05  &0.01 &M&5.30   &0.02   &198.10 &2.00   &0.300  &0.317  &$-$0.21\\
Gl876           &M4     &10.18  &0.01 &M&5.01   &0.02   &214.60 &0.20   &0.293  &0.334  &$+$0.03\\
Gl880           &M1.5   &8.70   &0.02 &M&4.52   &0.02   &145.30 &1.20   &0.569  &0.586  &$+$0.05\\
Gl896A          &M3.5   &\multicolumn{2}{c}{10.38}   &G&5.33   &0.02   &160.10 &2.80   &0.333  &0.387  &$+$0.08\\
Gl896B          &M4.5   &\multicolumn{2}{c}{12.40}   &G&6.26   &0.04   &160.10 &2.80   &0.186  &0.248  &$+$0.14\\
Gl908           &M1     &8.99   &0.01 &M&5.04   &0.02   &167.50 &1.50   &0.479  &0.421  &$-$0.52\\
\hline
\end{tabular}
\end{center}
$^\dagger$ M: Mermilliod, Mermilliod \& Hauck (\cite{mermilliod97}), G: Gliese \& Jareiss (\cite{gliese91}), W: Weis (\cite{weis91}).\\
$^\ddagger$ K-band photometry given by Leggett (\cite{leggett92}) and transformed to 2MASS system using the relation given by Carpenter (\cite{carpenter01}).\\
The mass estimates use the listed photometry and the M/L relations of 
Delfosse et al. (\cite{delfosse00}).\\
The K-band photometry originates from 2MASS and the parallaxes are adopted
from the compilation of Delfosse et al. (2005, in prep).
\end{table*}

\section{Metallicity of M-dwarf planet-host stars}
\label{sect:planets}

It is now well established that planet host stars are more metal-rich 
than the average solar neighbourhood population (Gonzalez \cite{gonzalez97}; 
Santos et al. \cite{santos01}, \cite{santos03}, \cite{santos04}). Santos et 
al. established that the planet frequency rises very steeply with stellar 
metallicity, at least for [Fe/H] $>$ 0. While only $\sim$3\% of the solar 
metallicity stars are orbited by a (detected) planet, this fraction 
increases to over 25\% for stars with [Fe/H] above +0.3.

One leading explanation for this dramatic dependency is that the probability
of planet formation increases non-linearly with the mass of {\em dust} in
a proto-planetary disk. M dwarfs, with presumably smaller disks and hence
smaller disk dust mass at a given metallicity, provide a potentially
critical test of that idea. This has up to now been hampered by both
small statistic, with only two M-dwarf planet hosts known to date,
and the lack of reliable metallicity estimates for those stars. 
Our calibration resolves the second of those difficulties, and
shows that \object{Gl 876} and \object{Gl 436}, the two known M-dwarf 
planet-host stars, both have closely solar metallicities ($-$0.03 dex 
and $+$0.02 dex, respectively). Those unremarkable metal abundances 
do not shed light on whether M dwarf planet hosts are preferentially
metal-rich or not. Larger samples will be needed for that, and
our calibration will be a useful tool when they become available.

\section{Metallicity distribution of M dwarfs}
\label{sect:9pc}

Equation~\ref{eq:calib} allows us to estimate the metallicity of any 
individual 
M dwarfs with V- and K-band photometry and a well determined parallax. Here 
we use it to evaluate the metallicity distribution 
of the Delfosse et al. (2005, in prep) sample of northern M dwarfs 
within 9.25 parsecs. This volume-limited sample is believed to be complete,
and is therefore representative of the solar neighbourhood.  We removed all 
unresolved binaries as well as the faintest stars which are outside the 
validity range
of the calibration (K $\in$ [4\,mag,\,7.5\,mag]). Table~\ref{table:feh9pc}
lists the 47 remaining stars with their estimated metallicity.
For comparison, we consider a sample of 1000 non-binary solar-type stars
from the CORALIE radial-velocity planet-search programme (Udry et al. 
\cite{udry00}). This sample of single F, G or K dwarfs is representative 
of the solar neighbourhood, and we estimate their metallicity using
the Santos et al. (\cite{santos02}) calibration of the area of the 
cross-correlation function between the stellar spectra and an appropriate 
template. We display the two distributions and 
their cumulative functions (Fig.~\ref{fig:distribution4}). The two
distribution have similar shapes, but with a $\sim\,$0.07~dex
shift of the M-dwarf distribution towards lower metallicities.
A Kolmogorov-Smirnov test gives an $\sim\,8$\% probability 
that the two 
samples are drawn from the same parent distribution. 
The significance of the offset is therefore modest, but if real is
in the expected direction. Since M-dwarfs have much longer lifetimes than 
the age of the universe, every M-dwarf that ever formed is still here for
us to see, while some of the oldest solar-type stars have evolved to white
dwarfs. M dwarfs are thus expected to be slightly older on average, and 
from the age-metallicity relation therefore slightly more metal-poor.

\begin{figure}
	\centering
	\includegraphics[width=0.45\textwidth]{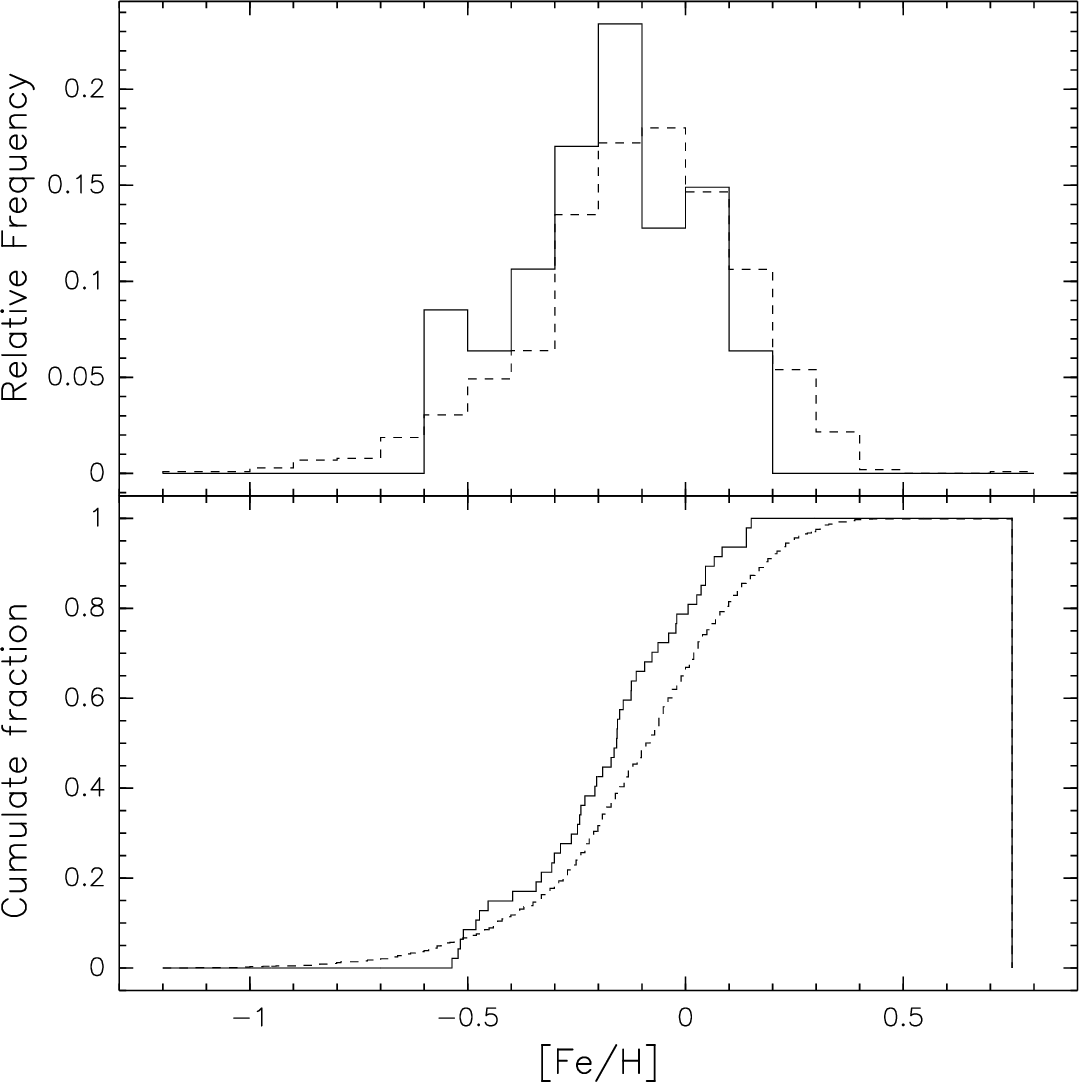}
   	\caption{{\it Upper panel :} M-dwarf metallicity distribution derived from Eq.~\ref{eq:calib} and, over-plotted in dashed line, the metallicity distribution of 1000 non-variable stars of our CORALIE radial-velocity planet-search programme. {\it Bottom panel :} Cumulative distributions of the same samples.
         	     }
         	\label{fig:distribution4}
\end{figure}

\section{Conclusions}
\label{conclusions}
We have determined the metallicities of 20 M dwarfs in wide-binary systems
that also contain an F, G or K star, under the simple assumption that the
two stars have the same composition. Where the parameter spaces overlap, 
our results are consistent with the direct analysis of M-dwarf spectra 
by Woolf \& Wallerstein (\cite{woolf05}). This provides a welcome
validation of both our assumptions of a common composition and of the WW05
analysis of complex M-dwarf spectra. The two datasets cover complementary
parameter ranges, and we join them to derive a photometric calibration of 
very low-mass star metallicities. The calibration is valid between 0.8
and 0.2~M$_\odot$, needs V- and K-band photometry and an accurate parallax, 
and provides metallicity estimates with $\sim$0.2~dex uncertainties. 
A 5\% parallax uncertainty results in an additional $\sim$0.2~dex metallicity
uncertainty, making the relation useful only within $\sim$50~pc.

We use the new metallicities to take a fresh look at the V-band 
{\it mass-luminosity} relation, and demonstrate that its intrinsic dispersion 
is indeed due to metallicity. We apply the new calibration to the two
known M dwarfs that host planets, \object{Gl 876} and \object{Gl 436},
and find  both of solar metallicity. Larger samples of M-dwarf planet
hosts will be needed to investigate whether they are preferentially
metal-rich, as are their solar-type counterparts. Finally, we estimate
metallicities for a volume-limited sample of 47 M dwarfs, and
compare its metallicity distribution to that of a much larger
sample of solar-type stars. The difference between the two
distributions is small, but if real might reflect slightly older
average ages for the long-lived M-dwarfs. In a forthcoming paper we 
will publish metallicities for a larger sample of M-dwarfs in binaries, 
observed from the southern hemisphere, and will attempt to derive a
purely spectrophotometric metallicity calibration.

\begin{acknowledgements}
We would like to acknoledge the anonymous referee for constructive
comments which led to an improved paper. This research
has made use of the SIMBAD database, operated at CDS, Strasbourg,
France. This publication makes use of data products from the Two Micron All
Sky Survey, which is a joint project of the University of Massachusetts and
the Infrared Processing and Analysis Center/California Institute of
Technology, funded by the National Aeronautics and Space Administration and
the National Science Foundation.
\end{acknowledgements}

\end{document}